\documentclass{llncs}
\usepackage[utf8]{inputenc}
\usepackage[T1]{fontenc}
\usepackage{amssymb} 
\usepackage{amsmath}
\usepackage{stmaryrd}
\usepackage{bbm}
\usepackage{graphicx}
\usepackage{paralist}
\usepackage{xcolor}

\usepackage{times}
\usepackage{soul}
\usepackage{subcaption}
\usepackage{url}
\usepackage{xspace}
\usepackage{enumitem}
\usepackage{multirow}
\usepackage{listings}
\usepackage{hyperref} 
\usepackage{tabularray}
\newcommand{\tool}{LLM4PR\xspace}

\begin{document}

\title{Towards Large Language Model Aided Program Refinement}

\author{
Yufan Cai\inst{1} \and
Zhe Hou\inst{2} \and
Xiaokun Luan\inst{3} \and
David Sanan\inst{4} \and\\
Yun Lin\inst{5} \and
Jun Sun\inst{6} \and
Jin Song Dong\inst{1}}
\authorrunning{Y. Cai et al.}

\institute{National University of Singapore \email{cai\_yufan@u.nus.edu} \email{dcsdjs@nus.edu.sg} \and Griffith University \email{z.hou@griffith.edu.au}
\and Peking University \email{luanxiaokun@pku.edu.cn} \and Singapore Institute of Technology \email{david.miguel@singaporetech.edu.sg} \and
Shanghai Jiaotong University \email{lin\_yun@sjtu.edu.cn} \and Singapore Management University \email{junsun@smu.edu.sg} }

\maketitle         
\begin{abstract}
Program refinement involves correctness-preserving transformations from formal abstract specification statements into executable programs. 
Traditional verification tool support for program refinement is highly interactive and lacks automation.
On the other hand, the emergence of large language models (LLMs) enables automatic code generation from informal natural language specifications. 
However, code generated by LLMs is often unreliable.
Moreover, the opaque procedure from specification to code provided by LLM is an uncontrolled black box.
We propose LLM4PR -- a tool that combines formal program refinement techniques with informal LLM-based methods to 
(1) transform the specification to pre- and post-conditions, 
(2) automatically build prompts based on refinement calculus, 
(3) interact with LLM to generate code, and finally,
(4) verify that the generated code satisfies the conditions of refinement conditions, thus guaranteeing the correctness of the code. 
We have implemented our tool with GPT4 and Coq and evaluated it on the HumanEval and EvalPlus datasets. 
\end{abstract}

\newcommand{\nonTerm}[1]{\ensuremath{\langle\mathit{#1}\rangle}}

\def\sectionautorefname{Section}
\def\subsectionautorefname{Section}
\def\algorithmautorefname{Algorithm}

\newcommand{\myParagraph}[1]{\par\medskip\noindent\emph{#1}~}

\definecolor{mGreen}{rgb}{0,0.6,0}
\definecolor{mGray}{rgb}{0.5,0.5,0.5}
\definecolor{mPurple}{rgb}{0.58,0,0.82}
\definecolor{backgroundColour}{rgb}{0.95,0.95,0.92}
\definecolor{backgroundColourOP}{rgb}{0.97,0.97,0.94}

\lstdefinestyle{CStyle}{
    backgroundcolor=\color{backgroundColour},   
    commentstyle=\color{mGreen},
    keywordstyle=\color{magenta},
    numberstyle=\tiny\color{mGray},
    stringstyle=\color{mPurple},
    basicstyle=\ttfamily,
    breakatwhitespace=false,         
    breaklines=true,                 
    captionpos=b,                    
    keepspaces=true,                 
    numbers=left,                    
    numbersep=5pt,                  
    showspaces=false,                
    showstringspaces=false,
    showtabs=false,                  
    tabsize=2,
    language=C
}

\lstdefinelanguage{ExtWhile}{
    morekeywords={skip, if, else, while, for, assert, assume, const, store, select, aggregate, int, verify},
    sensitive=false,
    morecomment=[l]{//},
    morecomment=[s]{/*}{*/},
}

\lstdefinestyle{ExtWhileStyle}{
    backgroundcolor=\color{backgroundColour},   
    commentstyle=\color{mGreen},
    keywordstyle=\color{magenta},
    numberstyle=\tiny\color{mGray},
    stringstyle=\color{mPurple},
    basicstyle=\ttfamily\footnotesize,
    breakatwhitespace=false,
    breaklines=true,
    captionpos=b,
    keepspaces=true,
    numbers=left,
    numbersep=5pt,
    showspaces=false,
    showstringspaces=false,
    showtabs=false,
    tabsize=2,
    language=ExtWhile
}

\lstdefinestyle{ExtWhileStyleOp}{
    backgroundcolor=\color{backgroundColourOP},   
    commentstyle=\color{mGreen},
    keywordstyle=\color{magenta},
    numberstyle=\tiny\color{mGray},
    stringstyle=\color{mPurple},
    basicstyle=\ttfamily\scriptsize,
    breakatwhitespace=false,
    breaklines=true,
    captionpos=b,
    keepspaces=true,
    numbers=left,
    numbersep=5pt,
    showspaces=false,
    showstringspaces=false,
    showtabs=false,
    tabsize=2,
    language=ExtWhile
}

\definecolor{nodeGray}{rgb}{0.96,0.96,0.96}
\definecolor{cornerGray}{rgb}{0.90,0.90,0.87}
\definecolor{beaublue}{rgb}{0.74, 0.83, 0.9}
\definecolor{mistyrose}{rgb}{1.0, 0.89, 0.88}


\newcommand{\assign}{\;\texttt{=}\;}
\newcommand{\eqeq}{\;\texttt{==}\;}
\newcommand{\semic}{\texttt{;}\;}
\newcommand{\add}{\;\texttt{+}\;}
\newcommand{\mult}{\;\texttt{*}\;}
\section{Introduction}
\label{sec:intro}
\paragraph{Background.}
Recently, AI-powered large language models (LLMs) have advanced rapidly in mathematics, reasoning, and programming~\cite{zhao2023survey,romera2023mathematical}.
Industrial products like GPT4~\cite{openai2023gpt4} and Copilot~\cite{copilot} greatly assist programmers in coding-related tasks.
In general, the programmer inputs a specification of the question in natural language, and then the LLM will generate the associated code, which basically \textit{translates} the natural language to the programming language. 
The end-to-end framework of deep learning models makes it possible to generate the intended program in a very flexible way.
Some studies, however, show that programmers usually find it hard to trust and debug the LLM-generated code \cite{vaithilingam2022expectation,ding2023static} as the generation procedure is opaque and out of control.
Past works like Code2Inv~\cite{Code2Inv,si2018learning} proposed an end-to-end learning framework to learn and generate a valid proof for a program by interacting with the proof checker.
With the emergence of LLM applications, recent works investigate methods that combine LLMs with formal verification techniques for generating program properties and invariants ~\cite{chakraborty2023ranking,anonymous2024lemur}. 
The recent works based on deep learning techniques usually adopt an end-to-end framework and rely on various informal heuristics like the chain of thoughts to control the reasoning of LLMs~\cite{wei2022chain}.
The verification procedure usually involves another LLM to check the output of the LLM in a debating-like procedure~\cite{zhang2023cumulative}. 

\paragraph{Challenges.}
While the above methods show the significant potential of LLMs in code generation and program verification, there remain questions in verifying and controlling the code generation procedure.
Besides, LLMs often generate unsound and insecure code that users would typically have no clue how to fix them. 
Building trust and interpretability in the code generation process is extremely important in practice since the generated code will be adopted in a large context and should be properly maintained.
As a complementary method, program refinement involves correctness-preserving transformations from formal specification statements into executable code.
However, the current transformation from specifications to code based on program refinement calculus is largely designed or even implemented by hand, which is costly and laborious ~\cite{ProgramFromSpecification,swierstra2016proposition,carrington1998program}.
Naturally, the manual transformation of program refinement always tends to be an ad-hoc procedure that is hard to generalize and apply in industry. 

\paragraph{Proposed Solution.}
In this work, we propose a mostly automated approach called \tool to combine the formal program refinement calculus with informal LLMs to refine the specification and generate verified code step by step automatically.
\tool also combines some automated theorem provers(ATPs) to verify the code and justify the choice of the refinement laws made by LLMs.
Our approach is complementary to LLMs, automated theorem provers, and traditional verification tools.
While the formal specification still requires manual effort for the initial input in the first step, it should not be a hurdle for the formal methods community and
is necessary since, otherwise, there is no correctness to speak of.
Besides, our LLM also facilitates the formalization procedure shown in the experiment.
To the best of our knowledge, \tool is the first framework that combines LLMs and program refinement techniques.

\paragraph{Motivating Example.}
We illustrate our motivation using a program for computing the \textit{square root of a real number}.
In \autoref{fig:motivate}, we show the code snippets generated by GPT4 and Copilot.
The LLMs can generate \textbf{almost} correct code.
However, these programs still contain some bugs.
Both upper two programs are wrong in the case $N < 1$ as $N*N < N$. 
Mathematically, the choice of the variable \texttt{high} as the upper bound of the square root of N should be larger than $N+\frac{1}{4}$ as $\forall N, (N+\frac{1}{4})^2 \ge N$.
We try to fix the GPT4 code with the prompt \textit{The upper bound is wrong for N less than 1}. 
However, the newly generated code still fails on several cases like sqrt(5) since the variable x goes to a fixed point but does not terminate the loop.
The final code (bottom right) with the formal constraints shows the conditions that should be obeyed for GPT4.
In contrast, our program refinement with the LLM will automatically generate prompts with constraints to generate the code and refine the specification shown in \autoref{case_study}.
Intuitively, we regard the LLMs as ``constraint solvers'', whose powerful extensibility and rich background knowledge shed light on the potential of automation for program refinement.
Our program refinement can passively ``assert'' constraints that help debugging and actively ``verify'' constraints that benefit code generation.

\begin{figure}[t!]
\centering
  \begin{minipage}{0.50\linewidth}
\begin{lstlisting}[style=ExtWhileStyle]
# GPT4
def find_x(N, e):
  low, high = 0, N
  while low <= high:
    mid = (low + high) / 2
    square = mid * mid
    if square < N and N < (mid + e) * (mid + e):
        return mid
    if square < N:
        low = mid + e
    else:
        high = mid - e
  return mid
\end{lstlisting}
  \end{minipage}
  \hfill
  \begin{minipage}{0.46\linewidth}
\begin{lstlisting}[style=ExtWhileStyle]
# Copilot
n = input()
n = float(n)
e = 0.0000001
low = 0
high = n
mid = (low + high) / 2
while abs(mid * mid - n) > e:
    if mid * mid > n:
        high = mid
    else:
        low = mid
    mid = (low + high) / 2
print(mid)
\end{lstlisting}
  \end{minipage}
  \hfill
  \begin{minipage}{0.50\linewidth}
\begin{lstlisting}[style=ExtWhileStyle]
# Repair when N is less than 1
def sqrt(N):
    x = N if N > 1 else 1
    while x * x > N:
        x = (x + N/x) / 2
    return x
\end{lstlisting}
\end{minipage}
  \hfill
  \begin{minipage}{0.46\linewidth}
\begin{lstlisting}[style=ExtWhileStyle]
# GPT4 + Constraints
x = N
assert x * x > N
while x * x > N:
    assert x != (x + N/x) / 2
    x = (x + N/x) / 2
\end{lstlisting}
\end{minipage}
\caption{Problematic code generated by GPT4 and Copilot for computing square root. The upper two programs are wrong in the case $N < 1$ due to the wrong upper bound initialization. The repaired code still fails in several cases when the variable x goes to the fixed point but does not terminate the loop due to the floating-point precision error. The last code contains the refinement constraints that point out the above problems.}
\label{fig:motivate}
\end{figure}

\paragraph{Contributions.}
The contributions of the paper are summarized below. 
\begin{enumerate}
\item A framework \tool for mostly automated program refinement with the LLMs, including a formal specification language $L_{spec}$, a programming language $L_{pl}$ associated with our program refinement calculus, and a verification strategy that verifies the outputs of LLM based on Coq and ATPs. 
\item A GPT4 variant fine-tuned with program refinement instructions and knowledge of our defined languages and laws.
\item A dataset of formal specifications and an evaluation benchmark based on the samples of the HumanEval and EvalPlus datasets.
\end{enumerate}

\section{Preliminaries}\label{Preliminary}
This section introduces the background knowledge of program refinement.
We mainly follow Morgan's notations in~\cite{ProgramFromSpecification}. 

\paragraph{Specification}
describes what a program is expected to do. 
In detail, a specification contains \textit{variants}, a \textit{precondition}, and a \textit{postcondition},
in the form
$$ variants: [precondition,~ postcondition].$$
Variants are the list of program variables,
the precondition describes the initial states, and the postcondition describes the final states of the program. 

\paragraph{Refinement}
of the specification is the relation between two expressions where one can \textit{solve} the other.
Formally, we have the following laws of refinement:

\begin{definition}[Strengthen Postcondition Law]
\label{def:strengthenPostconditon}
Let the precondition $pre$ and postcondition $post$ be any FOL formula,
if $ post' \Rrightarrow post$, then $x: [pre, post] \sqsubseteq x: [pre, post']$. 
\end{definition}

\begin{definition}[Weaken Precondition Law]
\label{def:weakenPreconditon}
Let the precondition $pre$ and postcondition $post$ be any FOL formula,
if $ pre \Rrightarrow pre'$, then $x: [pre, post] \sqsubseteq x: [pre', post]$. 
\end{definition}

The relation symbol $\sqsubseteq$ is called refinement.
For two formulae $A$ and $B$, $A$ entails $B$ ($A \Rrightarrow B$) means that in every state if $A$ is true then $B$ is true.

\paragraph{Skip} is a command where the final state of the program is the same as its initial state.
If the precondition of the specification entails the postcondition, it can be refined by skip.
\begin{definition}[Skip Law]
\label{def:skip}
If $\ pre \Rrightarrow post,\ then\ x: [pre, post] \sqsubseteq \textbf{skip}$. 
\end{definition}

\paragraph{Sequential Composition} refines a single specification to two others.
\begin{definition}[Sequential Composition Law]
\label{def:Sequence}
Let $mid$ be any formula except for $pre$ or $post$.
$ x: [pre, post] \sqsubseteq x:[pre, mid]\textbf{;}\ x:[mid, post]$. 
\end{definition}

\paragraph{Assignment} assigns the variant with new expressions.
We denote $post\langle x:=E \rangle$ as a new condition that assigns all occurrences of $x$ in $post$ by $E$.
If the precondition entails the new postcondition after the assignment, it can be refined by assignment.
\begin{definition}[Assignment Law]
\label{def:assignment}
Let $E$ be any $Expression$, $post\langle x:=E \rangle$ assigns every x in $post$ with $E$.
If $\ pre \Rrightarrow post\langle x:=E \rangle,\ then\ x: [pre, post] \sqsubseteq \textbf{x = E}$. 
\end{definition}

\paragraph{Alternation} is built by guarded branches.
\begin{definition}[Alternation Law]
\label{def:Alternation}
Let $GG$ be the disjunctive normal form of the guards $G_0, G_1, $
$, ..., G_i, ..., G_n$,
if $\ pre \Rrightarrow GG,\ then\ x: [pre, post] \sqsubseteq \textbf{if} \bigsqcup_i (G_i\ \textbf{then}\ x: [G_i\land pre,\ post]) $ where \textbf{if} $\bigsqcup_i G_i$ \ \textbf{then} means if $G_0$ then ... else if $G_i$ then ... . 
\end{definition}

\paragraph{Iteration.}
Iterations (while loops) are built by loop conditions, invariants, and variants.
An invariant $inv$ is a formula that if is true initially, stays true for each repetition. 
The variant $V$ of the iteration is chosen to guarantee the termination of the iteration.

\begin{definition}[Iteration Law]
\label{def:Iteration}
Let $Inv$, the invariant, be any formula; let $V$, the variant, be any integer-valued expression. Let $GG$ be the disjunctive normal form of the guards $G_0, G_1, ..., G_i, ..., G_n$
then $ x: [Inv,\ Inv \land \neg GG ] \sqsubseteq\ $ \textbf{while} $\bigsqcup_i(G_i\ \textbf{do}\ x: [Inv\land G_i,Inv\land (0\leq V<V_0)]) $ where $V_0$ is the initial value of V, \textbf{while} $\bigsqcup_i G_i$ \textbf{do} means while $G_0$ do ... else $G_i$ do ... else $G_n$ do.
\end{definition}

\paragraph{Expand.}
It expands the variant list by introducing another variant. 

\begin{definition}[Expand Law]
\label{def:Expand}
Let x be the origin variant and y be another variant and $y_0$ be the initial value of y, then 
$ x: [pre, post] = (x, y): [pre, post \land y = y_0]$
\end{definition}

\paragraph{Procedure.}
A procedure is declared by a name, some parameters, and a program.
\begin{definition}[Procedure]
\label{def:Procedure}
$ procedure~N~(param~V:T)~\triangleq Prog$.
\end{definition}

\begin{definition}[Procedure Value Specification]
\label{def:ProcedureRefine}
Given a procedure that refines \\
$ procedure~Proc~(param~f:T) \triangleq f: [pre, post]$, with $post$ containing no f. 
Let A be some expression, then 
$w: [pre\langle f:=A \rangle, post\langle f:=A \rangle] \sqsubseteq Proc(A)$
\end{definition}

\section{Formal Languages in Our Approach}\label{language}
We introduce our formal specification language $L_{spec}$ used to describe the specification and the programming language $L_{pl}$ for our generated code.
We further define the \textit{annotated programming language} for the program refinement procedure, which contains both $L_{spec}$ and $L_{pl}$. 
Formally, it is a tuple $(L_{spec}, L_{pl})$ that has two parts, one for each of the above languages, respectively. 
As these languages closely interacted with the LLMs, we target designing languages well understood and applied by LLMs.

\subsection{The Specification Language $L_{spec}$}
Our specification language $L_{spec}$ extends first-order logic (FOL) and is a subset of the language of Coq~\cite{barras1999coq}.
The LLMs are familiar with both FOL and Coq grammar.
We follow the standard syntax and semantics of FOL and highlight the following notations.

\paragraph{Variants and Constants.} 
We use lower case words like x, y, z to denote the \textit{variants} that will change in the refinement and upper case words like N, M to denote \textit{constants}. 
Both variants and constants should be typed. 
\paragraph{Relations and Functions.}
We use common \textit{relation} operators and \textit{function} operators in SMT, such as $<, =, +, -, *, /, Array[Int], Array[Int:Int]$.
\paragraph{Syntax.}
We define our specification based on the first-order theory and theory of arrays. 
The full syntax of $L_{spec}$ is given in \autoref{tab:syntax_spec},
where $\nonTerm{Specification}$ defines the specification that needs to be refined, 
$\nonTerm{Definition}$ defines the condition that the variants should satisfy, 
$\nonTerm{Params}$ defines the variants and constants. 
In the case of $\nonTerm{atom}$, $\nonTerm{Expr}_\texttt{0}$ denotes the previous value of the expression, 
$\nonTerm{Name}\texttt{[}\nonTerm{atom}\texttt{]}$ specifies the array selecting operation, 
and $\nonTerm{Name}[\nonTerm{atom}:\nonTerm{atom}]$ is used for array slicing operation.
The remainder of the syntax is standard FOL used in SMT solving.

\begin{table}[t!]
    \caption{Syntax of the specification language $L_{spec}$.}
    \label{tab:syntax_spec}
    \begin{align*}
        \nonTerm{Type} ~::=~~ & \texttt{bool} \mid \texttt{nat} \mid \texttt{Z} \mid \texttt{float} \mid \texttt{array} ~\nonTerm{Type}
        \\
        \nonTerm{Specification} ~::=~~ & \texttt{Precondition:} \nonTerm{Definition} \;\; \texttt{Postcondition:} \nonTerm{Definition}
        \\
        \nonTerm{Definition}~::=~~ & \nonTerm{Name} \nonTerm{Params} \;\texttt{:=}\; \nonTerm{Expr} \texttt{.}
        \\
        \nonTerm{Params} ~::=~~ & \;\texttt{(}\; \nonTerm{Name} \;\texttt{:}\; \nonTerm{Type} \;\texttt{)}\;
        \\ 
        \nonTerm{Expr} ~::=~~ & \nonTerm{Logit} \mid \nonTerm{Logit} \land \nonTerm{Expr}  \mid  \nonTerm{Logit} \lor \nonTerm{Expr}  \mid \neg \nonTerm{Expr} \mid \nonTerm{QExpr}
        \\ 
        \nonTerm{QExpr} ~::=~~ & \texttt{forall}|\texttt{exists} \; \nonTerm{Params} \; \nonTerm{Expr}
        \\
        \nonTerm{Logit} ~::=~~ & \nonTerm{Term} \mid
        \nonTerm{Term} \;\texttt{<}\; \nonTerm{Logit}  \mid
        \nonTerm{Term} \;\texttt{<=}\; \nonTerm{Logit} \mid
        \nonTerm{Term} \;\texttt{=}\; \nonTerm{Logit} 
        \\ \mid ~~~ &
        \nonTerm{Term} \;\texttt{>}\; \nonTerm{Logit} \mid
        \nonTerm{Term} \;\texttt{>=}\; \nonTerm{Logit} \mid
        \nonTerm{Term} \;\texttt{<>}\; \nonTerm{Logit} 
        \\ 
        \nonTerm{Term} ~::=~~ & \nonTerm{Factor} \mid
        \nonTerm{Factor} + \nonTerm{Term} \mid 
        \nonTerm{Factor} - \nonTerm{Term} 
        \\ 
        \nonTerm{Factor} ~::=~~ & \nonTerm{atom} \mid
        \nonTerm{atom} \mult \nonTerm{Factor} \mid 
        \nonTerm{atom} \ / \ \nonTerm{Factor}
        \\ 
        \nonTerm{atom} ~::=~~ & \nonTerm{Number} \mid \nonTerm{Variable} \mid  \nonTerm{Const} \mid \texttt{true} \mid \texttt{false}
        \\ \mid ~~~ &
        \;\texttt{-}\; \nonTerm{Expr} \mid \;\texttt{(}\; \nonTerm{Expr} \;\texttt{)}\;  \mid \nonTerm{Expr}_\texttt{0}
        \\ \mid ~~~ &
        \nonTerm{Name}\texttt{[}\nonTerm{atom}\texttt{]}
        \mid 
        \nonTerm{Name}[\nonTerm{atom}:\nonTerm{atom}]
        \\
    \end{align*}\vspace*{-10truemm}
\end{table}

\paragraph{Semantics.}
We follow the standard FOL semantics defined in Coq and only present the notable elements in \autoref{tab:semantics_spec}. 
Note that the theory of arrays is realized by relations and functions, similar to its treatment in the literature~\cite{claessen2002}.
\begin{table}[htb]
    \caption{The semantics of the specification language $L_{spec}$.}
    \label{tab:semantics_spec}
    \begin{align*}
        type\ T ~\iff~~ & A\ value\ set\ T \ \ \llbracket e_T \rrbracket \in T
        \\
        variants\ v : T ~\iff~~ & A\ value\ v \in T \ \ \llbracket v \rrbracket 
        \\
        constant\ c : T ~\iff~~ & A\ value\ c \in T \ \ \llbracket c \rrbracket = c
        \\
        functional\ operator\ f(T_1,T_2,...) : T ~\iff~~ & \llbracket f(a,b,...) \rrbracket = f(\llbracket a \rrbracket, \llbracket b \rrbracket,...)
        \\ 
        relational\ operator\ R(T_1,T_2,...) : Bool ~\iff~~ & \llbracket R(a,b,...) \rrbracket = R(\llbracket a \rrbracket, \llbracket b \rrbracket,...)
    \end{align*}\vspace*{-10truemm}
\end{table}

\subsection{The Program Language $L_{pl}$}
Our program language is mainly based on While language.
The language is kept simple to make it easier for the LLM to understand and generate.
The complete syntax of our program language is given in \autoref{tab:programs}.
Our programming language is imperative and has data types for booleans, natural numbers, integers, float, characters, and arrays. 
We include the extension of Array and Assert statements.
The array has a natural number index type and the reading, updating, and slicing operations.
To control the size and structure of programming, we also incorporate the use of procedures.
The procedure is declared by a name, some parameters, and an associated program follows \autoref{def:Procedure}.
The formal semantics follows the literature~\cite{köhl2021executable}. 

\begin{table}[tb]
    \caption{Syntax of the program language $L_{pl}$. }
    \label{tab:programs}

    \begin{align*}
        \nonTerm{Type} ~::=~~ & \texttt{bool} \mid \texttt{nat} \mid \texttt{int} \mid \texttt{float} \mid \texttt{char} \mid \texttt{array} \nonTerm{Type}
        \\
        \nonTerm{Expr} ~::=~~ & \nonTerm{Number} \mid \nonTerm{Name} \mid \texttt{true} \mid \texttt{false} \mid \nonTerm{Variable}
        \\ \mid ~~~ &
        \nonTerm{Expr} \eqeq \nonTerm{Expr} \mid
        \nonTerm{Expr} \;\texttt{<}\; \nonTerm{Expr}  \mid
        \nonTerm{Expr} \;\texttt{<=}\; \nonTerm{Expr}
        \\ \mid ~~~ &
        \nonTerm{Expr} \;\texttt{>}\; \nonTerm{Expr} \mid
        \nonTerm{Expr} \;\texttt{>=}\; \nonTerm{Expr} \mid
        \nonTerm{Expr} \;\texttt{!=}\; \nonTerm{Expr} 
        \\ \mid ~~~ &
        \nonTerm{Expr} \; \texttt{and}\; \nonTerm{Expr}  \mid
        \nonTerm{Expr} \; \texttt{or}\;   \nonTerm{Expr}  \mid
        \texttt{not}\nonTerm{Expr}
        \\ \mid ~~~ &
        \nonTerm{Expr} + \nonTerm{Expr} \mid 
        \nonTerm{Expr} - \nonTerm{Expr} 
        \\ \mid ~~~ &
        \nonTerm{Expr} \cdot \nonTerm{Expr} \mid 
        \nonTerm{Expr} \ / \ \nonTerm{Expr}
        \\ \mid ~~~ &
        \texttt{\nonTerm{Variable}\texttt{[}\nonTerm{Expr}\texttt{]}}
        \\ \mid ~~~ &
        \texttt{\nonTerm{Variable}[\nonTerm{Expr}:\nonTerm{Expr}]}
        \\
        \nonTerm{Prog} ~::=~~ & 
        \texttt{pass} \mid
        \nonTerm{Variable} \assign \nonTerm{Expr}
        \mid
        \nonTerm{Prog}\semic \nonTerm{Prog} 
        \\ \mid ~~~ &
        \texttt{while}~\texttt{(}\nonTerm{Expr}\texttt{):}~\nonTerm{Prog}
        \\ \mid ~~~ &
        \texttt{if}~\texttt{(}\nonTerm{Expr}\texttt{):}~\nonTerm{Prog}~\texttt{else}~\nonTerm{Prog}
        \\ \mid ~~~ &
        \texttt{assert} \nonTerm{Expr}
        \\ \mid ~~~ &
        \texttt{def}~\nonTerm{Name}~\texttt{(}\nonTerm{Name}~\texttt{:}~\nonTerm{Type}\texttt{)*:} \nonTerm{Prog}
        \\
    \end{align*}\vspace*{-10truemm}
\end{table}

\section{The Refinement Laws in Our Approach}\label{Law}
This section introduces our program refinement laws used for interaction with the LLMs. 
We aim to transform the refinement laws to facilitate both LLM interaction and ATPs verification.
Our defined refinement laws can be utilized by our LLM.

\paragraph{Skip.}
Another skip law gives the variant an initial value.
The new skip law utilizes the fact that the initial and final variables have the same value.

\begin{lemma}[Initialised Skip Law]
Let $x_0$ denote the initial value of variant $x$,
if$\ (x=x_0) \land P \Rrightarrow Q $, then the specification$ \ x: [P, Q] \sqsubseteq $ Skip. 
\end{lemma}
\begin{proof}
Use the skip law in \autoref{def:skip} as $P \Rrightarrow Q$.
\qed
\end{proof}

\paragraph{Seq.} 
We extend a new sequential composition law to flexibly divide one specification into two parts.

\begin{lemma}[Flexible Sequential Composition Law]
Let P, Q, A, B, C, D be some formulate, if$\ (P \Rrightarrow A) \land (B \Rrightarrow C) \land (Q \Rrightarrow D)$, then the specification$ \ x: [P, Q] \sqsubseteq x:[A, B]\textbf{;}\ x:[C, D]$. 
\end{lemma}
\begin{proof}
First, use the sequential composition law in \autoref{def:Sequence}, $ x: [P, Q] \sqsubseteq x:[P, B]; x:[B, Q]$.
Then refine the two parts with the weaken-precondition law in \autoref{def:weakenPreconditon}, $ x:[P, B] \sqsubseteq x:[A, B]; x:[B, Q] \sqsubseteq x:[C, Q] $. Finally refine the second part with the strengthen-postcondition law in \autoref{def:strengthenPostconditon}, $ x:[C, Q] \sqsubseteq x:[C, D] $.
\qed
\end{proof}

\paragraph{Assign.} 
We have two assignment laws.
The initialized assignment law utilizes the initial values of the variants to simplify the further proof for $\ pre \Rrightarrow post\langle x:=E \rangle$.
The following assignment law allows any assignment in its second half provided the changed variants. 

\begin{lemma}[Initialized Assignment Law]\label{def:initial_assignment}
Let $E$ be any $Expr$ in the programming language, $post\langle x:=E \rangle$ replaces every x in the formula $post$ with $E$.
If $ (x = x_0) \land (y = y_0) \land \ pre \Rrightarrow post\langle x:=E \rangle,\ then\ x, y: [pre, post] \sqsubseteq \textbf{x = E}$. 
\end{lemma}
\begin{proof}
Use the assignment law in \autoref{def:assignment} as $\ pre \Rrightarrow post\langle x:=E \rangle$.
\qed
\end{proof}

\begin{lemma}[Following Assignment Law]\label{def:follow_assignment}
Let $E$ be any $Expr$ in the programming language, $post\langle x:=E \rangle$ replaces every x in the formula $post$ with $E$.
$ x : [pre, post] \sqsubseteq x :[pre, post\langle x:=E \rangle]\ \textbf{;\ x = E}$. 
\end{lemma}
\begin{proof}
First use the sequential composition law, $ x: [pre, post] \sqsubseteq x:[pre, post\langle x:=E \rangle]\textbf{;}\ x:[post\langle x:=E \rangle, post]$.
Then refine the second part using the assignment law, $ x:[post\langle x:=E \rangle, post] \sqsubseteq \textbf{x = E}$. 
\qed
\end{proof}

\paragraph{Alternate.}
The if-else alternation law is a simplified version of the original one. 
Intuitively, it separates the specification into a case analysis.

\begin{lemma}[If-else Alternation Law]
Let P, Q, and G be some formulae, then the specification 
$ \ x: [P,\ Q] \sqsubseteq if\ (G)\ (x:[P \land G,\ Q])\ else\ (x:[P \land \neg G,\ Q])$.
\end{lemma}
\begin{proof}
As $ Pre \Rrightarrow G \lor \neg G$ based on the law of excluded middle, the lemma can be directly implied from the alternation law in \autoref{def:Alternation}.
\qed
\end{proof}

\paragraph{Iterate.}
We extend the origin iterative law to float numbers, which need to find an upper bound to guarantee the loop termination in finite time. 
The first new specification assigns the initial value to the invariant and the second specification preserves the invariant and changes the variant during the iteration until the negated guard condition holds.
In practice, based on the convergence of monotonic sequences of real numbers, we replace the existing condition with the monotonic and bounded condition given in \autoref{flexible_iteration}.
To avoid infinite loops, we add the assertion to check whether the expression $V$ decreases by at least the error bound of the floating-point precision.

\begin{lemma}[Initialised Iteration Law]\label{initial_iteration}
Let P, I, and G be some formulae, V be any variant expression, and i and M are positive integers, then the specification 
$ \ x: [P, I \land \neg G] \sqsubseteq x:[P, I]\ \textbf{;}\ while(G)\ do\ 
(x: [I \land G,\ I \land (\exists i < M,\ V_i \to \neg G)) ] $.
\end{lemma}
\begin{proof}
First, using the sequential composition law in \autoref{def:Sequence}, $ \ x: [P, I \land \neg G] \sqsubseteq x:[P, I]\textbf{;}\ x:[I, I \land \neg G]$.
Then refine the second part with the iteration law in \autoref{def:Iteration}.
Note that we replace the condition for integer-valued variants with any variant expression for scalability. 
To guarantee the termination of the iteration, a state of variant should exist to negate the guard condition after finite iterations. 
\qed
\end{proof}

\begin{lemma}[Flexible Iteration Law]\label{flexible_iteration}
Let P, I, and G be some formulae, V be any variant expression, then the specification 
$ \ x: [P, I \land \neg G] \sqsubseteq x:[P, I]\ \textbf{;}\ while(G)\ do\ 
(x: [I \land G,\ I \land V < V_0]; assert\ V \neq V_0) $.
\end{lemma}
\begin{proof}
First, follow the initialized Iteration Law. Then, note that the float precision error is $e$, then we have $\exists i = \lceil \frac{V_0}{e} \rceil < M, V < 0 \to \neg G$.
\qed
\end{proof}

\paragraph{Traverse.}
We build a traverse law to facilitate the problems related to the list with recurrence relation. 
The formula $P$ contains the variants $l$ and $i$, which can be equations that recursively define a sequence.
Note that the following refinement should preserve the invariant $P(l, i)$ and make progress to $P(l, i+1)$ following induction.

\begin{lemma}[Traverse Law]
Let l be the list of type T, natural numbers m and n denote the range, pre and P be some formula,
$ \ l: [pre, \forall i:nat \land m \le i < n \to P(l, i)] \sqsubseteq l, i:[pre, l[m]];\ i = m\ \textbf{;}\ while(i < n)\ do\ 
(l, i: [P(l, i), P(l, i+1)]\textbf{;}\ i=i+1) $.
\end{lemma}
\begin{proof}
First, using the expand law and sequential composition law in \autoref{def:Sequence}, $ \ l, i: [pre, l[i]\land i=m]; l, i: [l[i]\land i=m, l[i]\land i=n]$.
Then refine the second part with the initialised assignment law \autoref{def:initial_assignment} and iteration law in \autoref{def:Iteration}, we have $i = m\ \textbf{;}\ while(i < n)\ do\ (l,i: [P(l, i), P(l, i)\land 0\le n-i < n-i_0] $.
Finally, using the following assignment law in \autoref{def:follow_assignment} for the specification, $[P(l, i), P(l, i+1)\land 0\le n-(i+1) < n-i]\textbf{;}\ i=i+1$ and can be simplified to the target.
\qed
\end{proof}

\subsubsection{Semantics}
\begin{figure}[t!]
\centering
\includegraphics[width=1\textwidth]{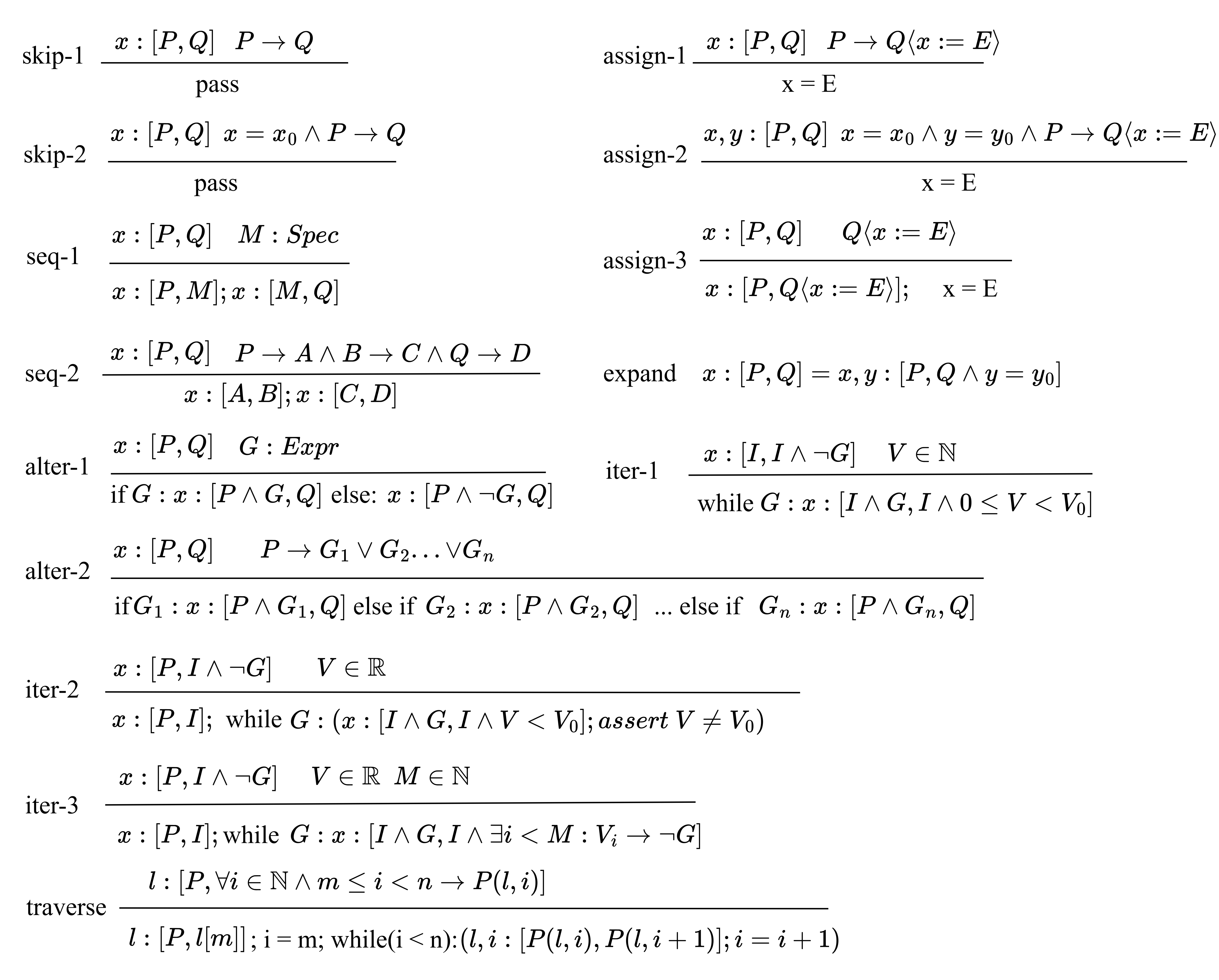}
\caption{Refinement laws semantics of the specification language and program language.}
\label{fig:semantics}
\end{figure}

The semantics shown in \autoref{fig:semantics} is defined according to the refinement laws proved above.

\section{\tool: Program Refinement with LLM}\label{Approach}
This section presents our approach that combines the above program refinement laws with the LLMs for automation.

\begin{figure}[t!]
\centering
\includegraphics[width=\textwidth]{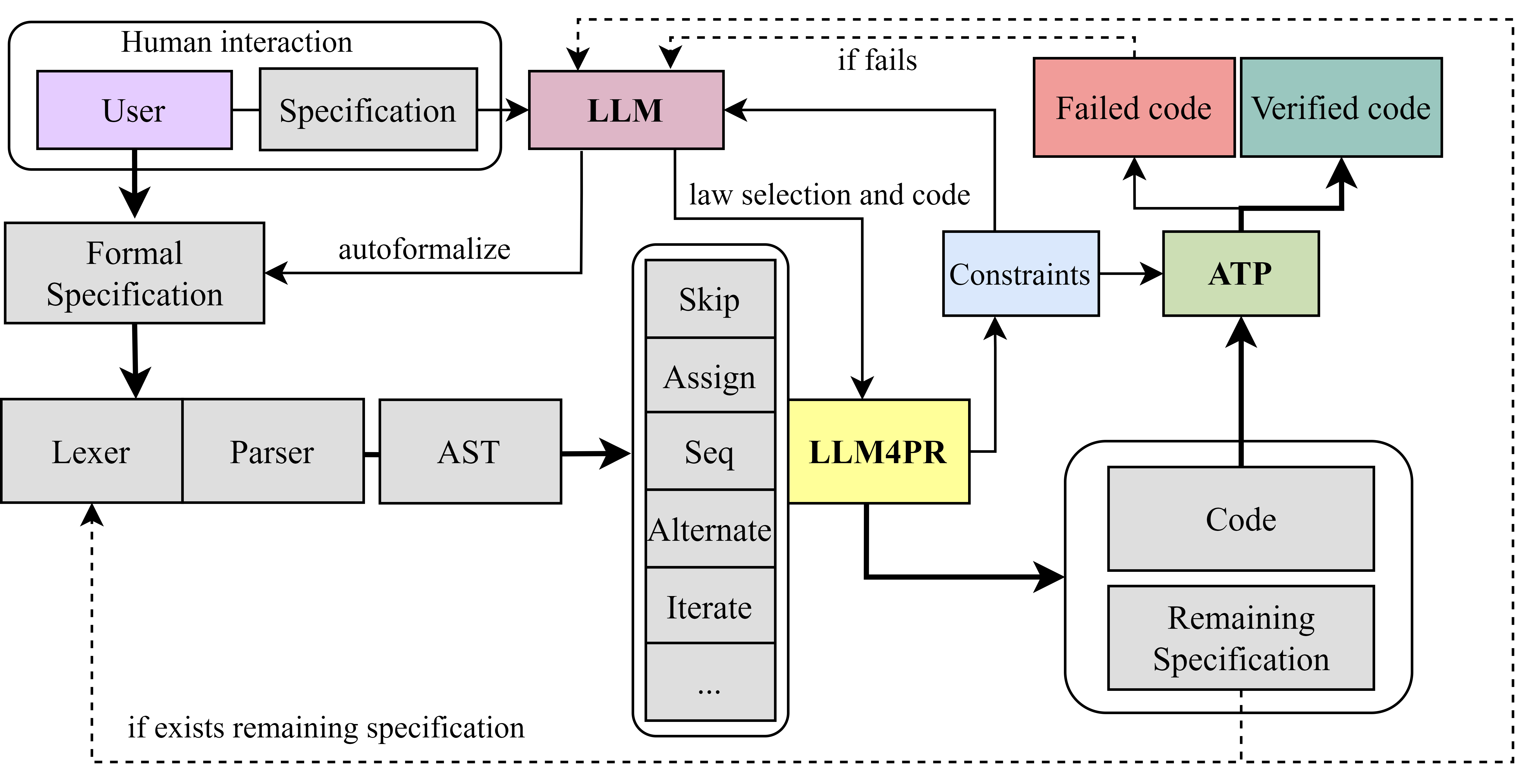}
\caption{Overview of \tool that combines LLMs and program refinement.}
\label{fig:approach}
\end{figure}

\paragraph{Overview.}
\autoref{fig:approach} shows an overview of our approach. 
The formal specification written in $L_{spec}$ will be first transformed to an abstract syntax tree and \tool will extract the conditions to input the LLM. 
The LLM will select a predefined law to refine the specification based on the description and constraints of the formal specification, and then generate the associate code with the law.
\tool correspondingly generates the proviso condition of the law and builds the verification scripts to justify the code generated by the LLM.
ATPs will try to automatically verify the scripts and output the success message or error message.
Based on the ATP result, the LLM will regenerate the code if failed, or the \tool will save the verified code and generate the new specification if succeeded.
If getting multiple times of failure, \tool will trace back to the last refinement step and specification and interact with the LLM to choose another law and generate the associated code.
\autoref{approchT} shows a summarization of the actions of LLM and \tool using the predefined six kinds of refinement laws.

\begin{table}[htb!]
\centering
\begin{tabular}{|l|c|p{9cm}|}
\hline
\textbf{Law} & \textbf{GPT4}         & \textbf{LLM4PR} \\ \hline
Skip         & -                     & verify $P \Rrightarrow Q$  \\ \hline
Sequence     & $M$                   & new spec $[P, M]; [M, Q]$                   \\ \hline
Assignment   & $x = Expr$            & verify $P \Rrightarrow Q\langle x := Expr \rangle $       \\ \hline
Alternation  & $G$                   & new spec \textit{if} ($G$) ($x: [P \land G, Q]$) else ($x: [P \land \neg G, Q]$)      \\ \hline
Iteration    & $G$                   & new spec $x:[P, I]$; while($G$) do($x:[ I \land G, I \land (\exists i < M,\ V_i \to \neg G)]$)       \\ \hline
Traverse     & -                     & new spec $l:[pre, l[m]];\ $i = m; while(i < n)\ do\ $(l, i: [P(l, i), P(l, i+1)];\ i=i+1) $          \\ \hline
\end{tabular}
\vspace{10px}
\caption{The schemes of specifications and conditions that the LLM and \tool generate for further verification in ATP. }
\label{approchT}
\end{table}

\paragraph{Actively Prompt.}
A \textit{prompt} for the LLM like GPT4 refers to the instruction given to the model to elicit a response.
The traditional design of prompts always follows some static templates like \textit{Program Refinement for the following specification}.
In this work, we regard the LLM as a constraint solver and actively build the prompt, including associated logical formulae for specifications step by step. 
These formulae contain the constraints that the output of LLM should satisfy.
Consequently, the prompts contain the constraints written in $L_{spec}$ and previous failure history if it exists. 
The LLM will select the refinement law and generate the associated code based on the given prompts. 
As each step has its generated specification, there is no need to add the previous refinement as history based on the congruence of \textit{Hoare Logic}. 

\paragraph{Passively Verify.}
After the LLM generates the choice of the law and the associated code, \tool will verify them using ATPs to justify whether the code satisfies the constraints based on the condition of the selected refinement law.
If the constraints can be satisfied, then \tool will apply the refinement law on the current specification and generate the new specification formally.
If the verification fails, the LLM will receive the failure message and the possible counterexamples and then try to generate another code. 
The trace-back process will repeat for a limited time. 
If it still fails, \tool will fall back to the last refinement step and the last specification.
The LLM will receive both the failure history and the specification.

\begin{figure}[t!]
\centering
\includegraphics[width=0.9\textwidth]{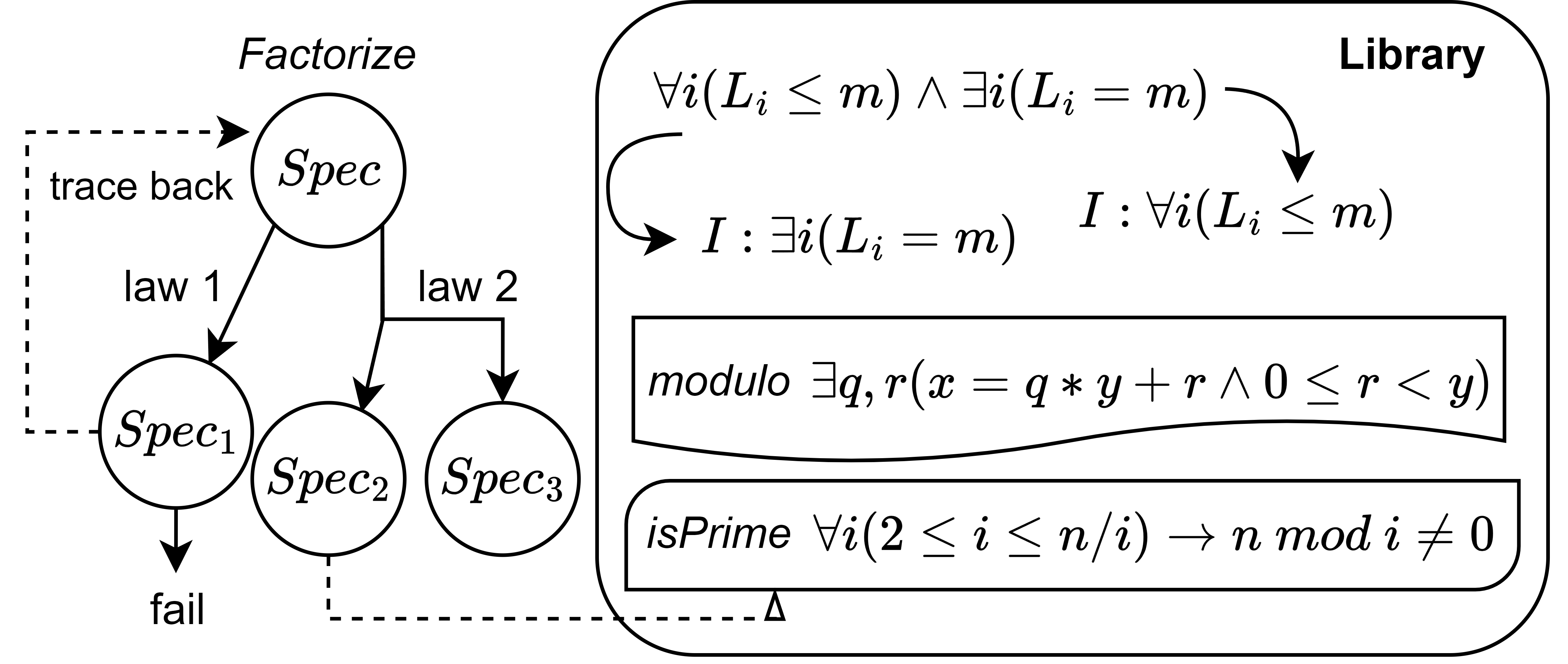}
\caption{The framework of the specification tree and the program refinement library. }
\label{fig:interaction}
\end{figure}

\paragraph{Refinement Procedures.}
The refinement procedure can be regarded as a specification tree, and the nodes are linked by the refinement laws. 
Each node has its specification and the possible refinement paths with the associated code.
We follow the procedure defined in \autoref{def:Procedure} and \tool will save the specification trees as a procedure in the refinement library for future reuse.
\autoref{fig:interaction} shows an overview of the specification tree and the refinement library.
The refinement procedures can be reused when meeting the new specifications that contain the same precondition and postcondition.
For example, the problem \textit{factorize} contains predicates like \textit{modulo} and \textit{isPrime} which can be referred to the library.

\paragraph{Retrieval Augmented LLM with Fine-tuning.}
We provide the LLM with the refinement procedures library as background knowledge so that the LLM can utilize the past knowledge with the retrieval augmented techniques.
We also customize the LLM for our program refinement task by crafting prompts and instructions based on the refinement laws mentioned above.
The LLM is fine-tuned with the examples in Morgan's book \cite{ProgramFromSpecification}, formal specification language $L_{spec}$, and our program language $L_{pl}$. 

\paragraph{Specification Formalization}
The first input of our \tool should be a formal specification that needs the user to formalize the requirement.
The LLMs can auto-formalize the specification to our $L_{spec}$, but it still needs the user to verify the correctness of the transformation from informal description to formal specification.
It should not be a hurdle for the formal methods community and is necessary since, otherwise, there is no correctness to speak of.

\section{Evaluation}\label{evaluation}
In this section, we first do the qualitative analysis of the detailed example to show the benefits of our approach and then the quantitative analysis of the most popular benchmark datasets compared to the state-of-the-art LLMs. 

\begin{figure}[ht!]
\centering
\begin{lstlisting}[escapechar=@,style=ExtWhileStyle]
// pre:  (N:float)(e: float) := N >= 0 /\ e > 0 
// post: (x:float)(y: float) := x*x <= N < y*y /\ y <= x+e
# LLM selects Sequential Composition Law: Part 1
// pre_1:= N >= 0 /\ e > 0 
// post_1:= x*x <= N < y*y
# LLM selects Assignment law
x = 0
y = N+1 
# verify pre_1 -> post_1(x := 0, y := N+1)
# Part 2
// pre_2:= x*x <= N < y*y
// post_2:= x*x <= N < y*y /\ y <= x+e
# LLM selects Iteration law: I(pre_2) G(~(y <= x+e))
while y > x+e:
  if y > x+e:
    // pre_2_1:= pre_2 /\ x+e < y 
    // post_2_1:= pre_2 /\ (...)
    # LLM selects Alternation law G((x+y)/2*(x+y)/2 > N)
    if (x+y)/2*(x+y)/2 > N:
      // pre_2_1_1:= pre_2_1 /\ (x+y)/2*(x+y)/2 > N
      // post_2_1_1:= post_2_1 /\ (...)
      y = (x+y)/2
      # verify pre_2_1_1 -> post_2_1_1(y := (x+y)/2)
    else:
      // pre_2_1_2:=  pre_2_1 /\ (x+y)/2*(x+y)/2 <= N
      // post_2_1_2:= post_2_1 /\ (...)
      x = (x+y)/2
      # verify pre_2_1_2 -> post_2_1_2(x := (x+y)/2)
\end{lstlisting}
    \caption{Example of a square root program in program refinement. The statements with \# are the explanation and the statements with // are the specifications.  }
    \label{fig:sqrt_example}
\end{figure}
\subsection{Case Study}\label{case_study}
We show how \tool deals with the motivating example of \autoref{sec:intro} in \autoref{fig:sqrt_example}. 
More examples are shown in \cite{LLM4PR}.
The statement tagged with \textit{\#} for code comments and the precondition and postcondition is the current specification. 
The verification statement is the proviso condition to apply the refinement law.
Note that we remove the condition of iteration termination check in (...) for a concise presentation.
In detail, the LLM first sequentially splits the original specification into two parts.
Informally, the first specification defines $x$, $y$ such that $x^2 \le N < y^2$, which can be implemented by assignment.
Note that the assignment of $y$ needs to satisfy the constraints in the postcondition of the specification that is $N < y^2$, eliminating the possibility of bugs of LLMs like $y = N$ in \autoref{fig:motivate}.
The second specification preserves the invariant $x^2 \le N < y^2$ and makes the variants $x$, $y$ closer until $x + e >= y$, which can be implemented with the iteration.
The invariant, the guard condition, and the variant can be extracted from the specification. 
The LLM reduces the distance between $x$ and $y$ to assign x or y with the mean of $x$ and $y$.
It uses alternation to add another constraint to strengthen the precondition and make it easier to conclude the postcondition.

Compared to the LLM-generated code, each refinement step can be verified as each has its associated specifications.
LLMs, on the other hand, are used to select the refinement law and generate associated code automatically based on the generated constraints.
The constraints are automatically built based on the choice of the law and the generated code in \tool.
When the refinement law is applied, the new specification will be generated based on the refinement laws with \tool.

\subsection{Experiments}
\subsubsection{Dataset and Implementations}
We choose the HumanEval dataset as the benchmark, which is widely used in evaluating code generation~\cite{openai2023gpt4,codex,wang2023codet5}.
To evaluate \tool with formal specifications, we transform 157 examples in the HumanEval dataset to a formal specification dataset, where 115 examples are correctly transformed by GPT4 and all formal specifications are manually checked.
Note that 7 examples can not be transformed to formal specifications.
Besides, to test the correctness and robustness of the generated code, we adopt the \emph{EvalPlus} \cite{evalplus} dataset with the same examples but average more than 80x the number of test cases. 
We choose GPT4 as the base model and fine-tune it with examples from Morgan's book \cite{ProgramFromSpecification} and then test \tool on the above dataset.

\subsubsection{Results}
\autoref{tab:result} shows the evaluation results.
We choose the state-of-the-art LLMs include LLama3~\cite{rozière2023code}, GPT-3.5, claude-3~\cite{enis2024llm} and GPT4 as our baselines.
The baselines' results follow the work \cite{evalplus}. 
To be fair, we add experiments that incorporate the formal specifications with the natural language descriptions to the GPT4.
With only the natural language descriptions as input, GPT4 shows the best overall performance of all the LLMs.
However, all the LLMs' performance decreases from HumanEval to EvalPlus because EvalPlus contains more challenging test cases and LLMs' generated code may have some bugs that can not pass the extra test cases.
In contrast, \tool's performance is consistent between HumanEval and EvalPlus as the code is verified with guaranteed correctness.
Theoretically, our generated code can be regarded as canonical solutions regardless of the number of test cases.
Interestingly, incorporating formal specifications also enhances the GPT4 since formal specifications contain useful constraints and information.
Overall, the \tool shows better performance and generates more robust code compared to the LLMs. 
\begin{table}[t!]
\centering
\bgroup
\def\arraystretch{1.2}
\setlength\tabcolsep{5px}
\begin{tabular}{|l|ccccc|c|}
\hline
Model               & \multicolumn{1}{c|}{Llama3} & \multicolumn{1}{c|}{GPT-3.5} & \multicolumn{1}{c|}{claude-3} & \multicolumn{2}{c|}{GPT4}         & LLM4PR \\ \hline
Input Specification & \multicolumn{1}{c|}{NL}     & \multicolumn{1}{c|}{NL}      & \multicolumn{1}{c|}{NL}       & \multicolumn{1}{c|}{NL}  & NL+ FS & FS     \\ \hline
Total \# Programs             & \multicolumn{4}{c|}{164}                                                                                              & 157    & 157    \\ \hline
HumanEval Passed          & \multicolumn{1}{c|}{125}    & \multicolumn{1}{c|}{126}     & \multicolumn{1}{c|}{136}      & \multicolumn{1}{c|}{145} & 148    & \textbf{150}    \\ \hline
EvalPlus Passed           & \multicolumn{1}{c|}{116}    & \multicolumn{1}{c|}{116}     & \multicolumn{1}{c|}{126}      & \multicolumn{1}{c|}{128} & 142    & \textbf{150}    \\ \hline

\end{tabular}
\egroup
\vspace{10px}
\caption{A comparison of \tool and popular LLMs on the number of generated programs that passed the test cases in HumanEval and EvalPlus datasets. NL means natural language inputs and FS means formal specification inputs.}
\label{tab:result}
\end{table}

\subsection{Limitations}
First, the capability of \tool largely depends on the capabilities of LLMs and ATPs.
To remedy the limitation, users can be involved in the procedure of program refinement by selecting the law, building, and checking the proof.
Second, if a refinement lacks proof of the loop termination in the iterative law, we still consider it as \emph{partially correct}.
We create more iteration laws and the traverse law in \autoref{Law} to help avoid the termination condition, as proving termination is a hard and generally undecidable problem.
Third, \tool is designed to guide LLM in generating more robust code, not for analyzing problems and building specifications, where the latter still requires human input.
However, it should not discount our approach since a definition of correctness is necessary for verified code.
\section{Related Work}

\paragraph{Program Refinement.}
\cite{backIntroduction}, \cite{ProgramFromSpecification}, \cite{back1990refinement} defines a formal method to build a program from its specification.
It mainly focuses on the correctness of a given specification and refinement of a program while preserving its correctness.
Some works propose a formalization of the refinement calculus in interactive theorem provers such as \cite{foster2020differential} for Isabelle and \cite{Refinecoq} for Coq \cite{barras1999coq}.
Recent works utilize refinement calculus on different applications including ~\cite{foster2020differential,dragomir2020refinement,elrakaiby2022care}.

\paragraph{Theorem Proving.}
There are two main types of tools for theorem proving: Interactive Theorem Provers (ITPs) and Automated Theorem Provers (ATPs) \cite{nawaz2019survey}.
ITPs, also known as proof assistants, interact with humans in the process of proof building and development, like Isabelle \cite{paulson1994isabelle}, Coq \cite{barras1999coq}, Lean \cite{de2015lean}.
ATPs prove the goals automatically, including E-prover \cite{Eprover}, cvc4 \cite{barrett2011cvc4}, vampire \cite{kovacs2013vampire} and Z3 \cite{de2008z3}.
Some ITPs also incorporate automated provers like Isabelle with Sledgehammer \cite{bohme2010sledgehammer} and Coq with Coqhammer \cite{czajka2018coqhammer}.

\paragraph{Formal methods with LLM.}
Recent research on generating formal mathematical proofs utilizes machine learning techniques for proof search and premise selection.
Existing works like GPT-f \cite{polu2020generative}, PACT \cite{han2021proof}, Expert Iteration \cite{polu2022formal} use LLMs to generate actions, and the search engine tries to find possible correct steps using the actions provided by the model.
Some works including HTPS \cite{lample2022hypertree}, and DT-Solver\cite{wang2023dt} enhance the search engine by machine learning techniques.
Thor \cite{jiang2022thor} uses the neural policy models incorporating ATPs to prove the theorems.
LeanDojo \cite{yang2023leandojo} enables interaction with the proof environment Lean \cite{de2015lean}.
It extracts fine-grained annotations of premises in proofs from Lean, providing valuable data for premise selection. 

\paragraph{Verification with LLM.}
One of the key challenges of LLMs is their tendency to "hallucinate", which refers to generating information that is not just incorrect but often fabricated specious text.
\cite{Dehallucinate} sketches a self-monitoring and iterative prompting way that uses formal methods to detect the hallucination and steer the LLM to the correct specification.
\cite{charalambous2023new} builds the specialized prompt based on counterexamples provided by model checking and conducts code debugging and repairing based on LLMs.

\section{Conclusion}
We have presented a tool \tool for \emph{automated} generation of \emph{verified} code using LLMs, Coq and ATPs.
We formally transform the specifications into code based on our refinement laws and LLMs. 
Our approach also extends the formal refinement calculus and builds active prompts to the informal LLMs.
Finally, \tool uses the ATPs to verify the refinement condition and the code based on the precondition and postcondition of the specification. 
Our experiments show that our method can generate more robust and correct code compared to the state-of-the-art LLMs.

\bibliographystyle{splncs04}
\bibliography{mybibliography}
\end{document}